\documentstyle[aps,psfig]{revtex}
\newcommand{\be}{\begin{equation}}
\newcommand{\ee}{\end{equation}}
\newcommand{\ben}{\begin{eqnarray}}
\newcommand{\een}{\end{eqnarray}} 
\def\pls{\partial\!\!\!/}
\def\bs{b\!\!\!/}
\def\ps{p\!\!\!/}
\def\As{A\!\!\!/}

\def\m{\mu}

\begin{document}

\twocolumn[\hsize\textwidth\columnwidth\hsize\csname 
@twocolumnfalse\endcsname

\title{RADIATIVELY INDUCED LORENTZ AND CPT VIOLATION \\
 IN QED AT FINITE TEMPERATURE}

\author{J. R. S. Nascimento\footnote{e-mail: jroberto@fisica.ufpb.br}
and R. F. Ribeiro\footnote{e-mail: rfreire@fisica.ufpb.br}}

\address{Departamento de F\'\i sica,
Universidade Federal da Para\'\i ba, Caixa Postal 5008, 58051-970 Jo\~ao Pessoa, Para\'\i ba, Brazil}

\author{N. F. Svaiter\footnote{e-mail: nfuxsvai@lafex.cbpf.br}}

\address{Centro Brasileiro de Pesquisa F\'\i sicas-CBPF, Rua Dr. Xavier 
Sigaud 150, Rio de Janeiro, RJ 22290-180 Brazil}

\date{\today}

\maketitle

\begin{abstract}

In this paper we evaluate the induced Lorentz and CPT violating Chern-Simons term in the QED action
at finite temperature. We do this using the method of derivative expansion of fermion 
determinants. Also, we use the imaginary-time formalism to calculate the temperature dependence of the Chern-Simons term.
\end{abstract}
\vskip2pc]

Recently, the problem of the radiatively induced Chern-Simons term in 
3+1 dimensions in quantum field theory has been addressed in several 
of papers. Particularly, this term has been  of much  interest in 
study of models that violate Lorentz and CPT symmetry\cite{JAK,MP,D1,CH}. 
In Quantum Electrodynamics, it is known that the Lorentz and CPT symmetry 
is destroyed by adding the Chern-Simons term to the Lagrangian density, 
${\cal L_{CS}} = \frac{1}{2} k_{\mu}\epsilon^{\mu\alpha\beta\gamma}F_{\alpha\beta}A_{\gamma}$ ~\cite{D1,Ja,D}, where $k_{\mu}$ is a constant 4-vector. 
The Chern-Simons term can be induced by CPT and Lorentz violating axial 
quantity, $b^{\mu}\gamma_{\mu}\gamma_{5}$, in the Lagrangian 
for massive fermions. Therefore, we get the modified quantum 
electrodynamics  which predicts birefringence of light in vacuum and  
observation of distant galaxies puts a stringent bound on 
$k_{\mu}$~\cite{Ja,D,CO,Go}.

The purpose of this paper is to compute the induced Chern-Simons term and 
analyze the behavior of the coefficient of this term
when we are taking temperature into account. To do this,
we use derivative expansion method of the fermion 
determinant\cite{DAS}  and the imaginary-time formalism 
developed by Matsubara.

Let us consider a modified QED theory described by Lagrangian density
\be
\label{lag}
{\cal L} = \bar\psi\left[i\pls - m - \gamma_5\bs - e\As\right]\psi, 
\ee
where $b_{\mu}$ is a constant 4-vector coupled to axial current and $A_{\mu}$ is 
the external field . 
We are using natural units, $c=\hbar=1$,  and the Dirac representation for 
Dirac matrices.  

The corresponding generating functional is
\be
\label{fg}
Z[A]=\int D\bar\psi(x)D\psi(x)\exp\left[i\int{\cal L}\;\;d^4x\right]
\ee 
We substitute Eq.(\ref{lag}) into Eq.(\ref{fg}) and integrate over the fermion 
field  we obtain
\be
\label{Seff}
Z[A] = Det(i\pls - m - \gamma_5\bs - e\As)= \exp[iS_{eff}[A]].
\ee
No Legendre 
transformation is required to go from the 
connected vacuum functional to the effective action. Thus, The effective 
action can be written as 
\be
\label{se}
S_{eff}[A] = -iTr\ln [i\pls - m - \gamma_5\bs - e\As].
\ee 

Let us write the trace in Eq.(\ref{se}) in the following equivalent form,
\ben
\label{id}
Tr\ln [i\pls - m - \gamma_5\bs - e\As] = Tr\ln [i\pls - m - \gamma_5\bs ] + F[A]
\een
where
\be
F[A] = \int_0^1 dz Tr\left[\frac{1}{i\pls - m - \gamma_5\bs - 
ze\As(x)}e\As(x)\right].
\ee
As $\partial_{\mu}$ and $A_{\mu}(x)$ do not commute, and to perform 
the momentum space integration of the second term in Eq.(\ref{id}), we use 
the notation $i\pls \rightarrow \ps$ and 
$\As(x) \rightarrow \As(x-i\frac{\partial}{\partial p})$. Then the 
Eq.(\ref{se}) can be written as 
\be
S_{eff}[A] = S_{eff}^{(0)}[A] + S_{effe}^{(1)}[A],
\ee
where
\ben
\label{seff}
S_{eff}^{(0)}[A] =-i Tr\ln [i\pls - m - \gamma_5\bs ]
\een
and
\ben
S_{eff}^{(1)}[A] &=& i \int_0^1 dz 
\int\frac{d^4p}{(2\pi)^4}\\ \nonumber
&\times& tr\left[\frac{1}{\ps - m - \gamma_5\bs - ze\As(x-i\frac{\partial}{\partial p_\m})}\As(x)\right].
\een

The Eq.(\ref{seff}) has been analyzed in detail in  
Ref.\cite{D}. Here, we study the second term of $S_{eff}^{(1)}[A]$, and we are 
keeping only first order derivative terms which are linear in $\bs$ and 
quadratic in $\As$. Using the operator expansion
$$
\frac{1}{A-B} = \frac{1}{A} + \frac{1}{A}B \frac{1}{A} + \frac{1}{A}B \frac{1}{A}B \frac{1}{A} + \cdots 
$$
we can write Eq.(\ref{seff}) as
\ben
\label{seff1}
S_{eff}^{(1)}[A] &=& -\frac{e^2}{2}\int d^4 x \int \frac{d^4 p}{(2 \pi)^4} \nonumber \\ 
 &\times& {\rm tr} [\frac{1}{\ps - m}i\partial_{\m}\As\frac{\partial}{\partial p_{\m}}\frac{1}{\ps - m}\gamma_5\bs\frac{1}{\ps - m}\As \nonumber \\ 
&+& \frac{1}{\ps - m}\gamma_5\bs \frac{1}{\ps - m}i\partial_{\m}\As\frac{\partial}{\partial p_{\m}}\frac{1}{\ps - m}\As ]
\een
after use the relation
$$
\frac{\partial}{\partial p_\mu}\frac{1}{\ps -m} = -\frac{1}{\ps -m}\gamma^{\mu}\frac{1}{\ps-m}.
$$
As can be seen by power counting the momentum integral has terms which 
diverge logarithmically, and to regularize the expression, we use the 
dimensional regularization method. Carrying out the tr of the $\gamma$ 
matrices Eq.(\ref{seff1}) takes the form
\be
\label{seff2}
S_{eff}^{(1)}[A] =  -\frac{e^2}{2}\int d^4\, x\int \frac{d^4\, p}{(2 \pi)^4} \left[\frac{N}{(p^2 - m^2)^4}\right],
\ee  
where $N$ is given by
\ben
\label{numer}
N &=& -4i(p^2 - m^2)\left[\epsilon^{\alpha\beta\m\sigma}\left(3m^2+p^2\right)-4\epsilon^{\alpha\beta\m\nu}
p_{\nu} p^{\sigma}\right] \\ \nonumber 
&\times&b_\sigma \partial_{\mu}A_\alpha A_\beta.
\een
Lorentz invariance ensures that 
\be
\label{re}
\int\frac{d^Dq}{(2\pi)^D}q_{\mu}q_{\nu}f(q^2)=\frac{g_{\mu\nu}}{D}\int\frac{d^Dq}{(2\pi)^D}q^2f(q^2)
\ee
where $D$ is the dimension of the space time. So, N can be written as
\be
N = -12m^2i(p^2 - m^2)\epsilon^{\alpha\beta\m\sigma}
b_\sigma \partial_{\mu}A_\alpha A_\beta.
\ee

Observe that the terms that contain $p^2$ and $p_{\nu}p^{\sigma}$ in 
Eq.(\ref{numer}) cancel when we use the Eq.(\ref{re}) with $D=4$. 
In this way, the logarithmical divergent terms in 
 Eq.(\ref{seff2}) disappear, so that we can rewrite the effective action as 
\ben
\label{seff3}
S_{eff}^{(1)}[A]& =&\left[ {6im^2 e^2}\int \frac{d^4\, p}{(2 \pi)^4} \left(\frac{1}{(p^2 - m^2)^3}\right)\,\right] \\ \nonumber 
& \times & \epsilon^{\alpha\beta\m\sigma}{b_{\sigma}}\int d^4\,x \partial_{\mu}A_\alpha A_\beta,
\een
Note that Eq.(\ref{seff3}) is finite by power counting.


From now on, we are interested in 
the induced Chern-Simons term taking 
temperature into account.(For a review see \cite{DJ}.) To do this, 
it is convenient to work in Euclidean 
space then Eq.(\ref{seff3}) can be written as
\ben
\label{seff4}
S_{eff}^{(1)}[A] &=& \left[ {6m^2 e^2}\int \frac{d^4\, p_E}{(2 \pi)^4} \left(\frac{1}{(p_E^2 + m^2)^3}\right)\,\right]\\ \nonumber  &\times& \epsilon^{\alpha\beta\m\sigma}{b_{\sigma}}\int(-i) d^4\,x_E \partial_{\mu}A_\alpha A_\beta
\een  
where for Euclidean momentum $p_{\mu}$, we define the real variable $p_4$ by 
$p_0=ip_4$ Thus, $p^2=-p_E^2$, and
$p_E^2= {\bf p^2} + p_4^2$, the standard Euclidean square of a vector. Also, we 
have considered $d^4p=id^4p_E$ , $x_0= -ix_4$,  and $d^4x=-id^4x_E$.

If we assume that the system is in thermal equilibrium with a reservoir at temperature $\beta^{-1}$ we may use the Matsubara 
formalism. In this case we have to perform the replacements $w \to w_n = (n + 1/2)2\pi/{\beta}$ and 
$(1/2\pi)\int dp_E^0=1/\beta\sum_n$. 
Then Eq.(\ref{seff4}) takes the form
\ben
\label{seff5}
S_{eff}^{(1)}[A] = -i6e^2 f(m^2, \beta) \epsilon^{\alpha\beta\m\sigma}{b_{\sigma}}\int d^4\,x_E \partial_{\mu}A_\alpha A_\beta
\een
where $f(m^2, \beta)$ is given by
\ben
\label{f}
f(m^2, \beta)=\frac{m^2}{\beta}\int\;\frac{d^3{\bf p} }{(2\pi)^3}\sum_{n=-\infty}^{\infty} \left(\frac{1}{({\bf p}^2+w_n^2 + m^2)^3}\right).
\een
First of all, we performing the sum operation in Eq(\ref{f}).  As result, we obtain the polynomial expression of third order in $\tanh(x)$. 
The follow in  step is substitute $\tanh(x) \to 1 - 2{\cal F_D}$ and ${\cal F_D}$ is given by
\be
{\cal F_D}= \frac{1}{\exp(\beta\sqrt{{\bf p^2} + m^2 }) +1}.
\ee
The function  $f(m^2, \beta)$ takes the form
\ben
\label{f2}
f(m^2, \beta) &=& \int\;\frac{d^3{\bf p} }{(2\pi)^3}\left[{\cal A}({\bf p^2}, m^2, \beta)\right. \\ \nonumber& + & \left.
{\cal B}({\bf p^2}, m^2, \beta) \beta 
+ {\cal C}({\bf p^2}, m^2, \beta)\beta^2 \right]
\een
The terms 
${\cal A}({\bf p^2}, m^2, \beta)$, ${\cal B}({\bf p^2}, m^2, \beta)$ 
and  ${\cal C}({\bf p^2}, m^2, \beta)$ are
\ben
\label{A1}
{\cal A}({\bf p^2}, m^2, \beta) = \frac{3}{16}\frac{m^2}{(p^2 + m^2)^{5/2}}\left[ 1 - 2{\cal F_D}\right],
\een
\be
\label{B1}
{\cal B}({\bf p^2}, m^2, \beta) = -\frac{3}{8}\frac{m^2}{(p^2 + m^2)^{2}}{\cal F_D}\left[1-{\cal F_D}\right],
\ee
and
\ben
\label{C1}
{\cal C}({\bf p^2}, m^2, \beta) = -\frac{1}{8}\frac{m^2}{(p^2 + m^2)^{3/2}}{\cal F_D}\left[1-3{\cal F_D}+2{\cal F_D}\right]
\een
observe that at zero temperature  the term ${\cal F_D} \rightarrow 0$ and we have
$${\cal A}({\bf p^2}, m^2, \beta \rightarrow \infty) \rightarrow \frac{3}{16}\frac{m^2}{(p^2 + m^2)^{5/2}},$$ while the other terms  
${\cal B}({\bf p^2}, m^2, \beta\rightarrow \infty)\rightarrow 0$ and 
${\cal C}({\bf p^2}, m^2, \beta\rightarrow \infty) \rightarrow 0$. 
So, at zero temperature we get
\be
\label{f21}
f(m^2, \beta) =\frac{3}{16} \int\;\frac{d^3{\bf p}}{(2\pi)^3}\frac{m^2}{(p^2 + m^2)^{5/2}}.
\ee 
To evaluate the momentum integration we use the relation
\ben
\label{for}
\int\frac{d^Dp}{{(2\pi)}^D} \frac{1}{(p^2 + a^2)^s} &=& \frac{1}{(2\pi)^D}\frac{\pi^{D/2}}{\Gamma(s)}\Gamma(s-D/2)\\ \nonumber &\times&
\frac{1}{{(a^2)}^{(s-D/2)}},
\een
so that the Eq.(\ref{f21}) becomes
\be
\label{f22}
f(m^2, \beta) = \frac{1}{32\pi^2}.
\ee
We use this result into Eq.(\ref{seff5}) to obtain
\be
\label{T0}
S_{eff}^{(1)}[A]=\frac{3e^2}{16\pi^2}\epsilon^{\alpha\beta\mu\sigma}{b_{\sigma}}\int(-i) d^4\,x \partial_{\mu}A_\alpha A_\beta.
\ee
This is the Chern-Simons term induced to zero temperature. Observe that 
this result is finite and is consistent with the computation 
of the vacuum polarization diagram with fermion propagator 
$S_F=(\ps - m -\gamma_5\bs)$ made in the Ref.\cite{JAK,MP}.
 
It is straightforward to verify that in the limit of high temperature, $\beta \rightarrow 0$, the term ${\cal F_D} \rightarrow \frac{1}{2}$.
Then, ${\cal A}({\bf p^2}, m^2, \beta\rightarrow 0) \rightarrow 0$ and 
the other terms  ${\cal B}({\bf p^2}, m^2, \beta\rightarrow 0)$
and  ${\cal C}({\bf p^2}, m^2, \beta\rightarrow 0)$ are different of zero 
and independent of $\beta$. On the other hand, as we can see
in Eq.(\ref{f2}) these terms are multiplied by $\beta$ and $\beta^2$ 
respectively in the numerator thus $f(m^2, \beta \rightarrow 0)$ 
is zero.
 
Let us summarize our results:
The coefficient of the induced Chern-Simons term behaves monotonically descendent
with $\beta(1/T)$, such that when $\beta \rightarrow \infty$ $(T \rightarrow 0$
we reproduce the same result obtained by Jackiw and Kostelecky \cite{JAK}. In the 
limit of $\beta \rightarrow 0$$(T\rightarrow \infty)$ the coefficient goes to zero.
Therefore, Lorentz and CPT symmetries are restored only when the temperature goes to
infinity.  

\acknowledgments 
The authors would like to thanks D. Bazeia for 
reading the manuscript. and F. Brandt
and M. Gomes for many suggestions.

\end{document}